\documentclass[pra,twocolumn,showpacs,groupedaddress]{revtex4}
\usepackage{amssymb}
\usepackage[dvips]{graphicx}
\usepackage{bm}
\usepackage{amsmath}

\begin{document}

\title{Two-polariton bound states in the
 Jaynes-Cummings-Hubbard Model}
\author{M. T. C. Wong and C. K. Law}
\affiliation{Department of Physics and Institute of Theoretical
Physics, The Chinese University of Hong Kong, Shatin, Hong Kong
SAR, China}

\date{\today}
\begin{abstract}
We examine the eigenstates of the one-dimensional
Jaynes-Cummings-Hubbard (JCH) model in the two-excitation subspace.
We discover that two-excitation bound states emerge when the ratio
of vacuum Rabi frequency to the tunneling rate between cavities
exceeds a critical value. We determine the critical ratio as a
function of the quasi-momentum quantum number, and indicate that the
bound states carry a strong correlation in which the two polaritons
appear to be spatially confined together.
\end{abstract}

 \maketitle

The investigation of quantum electrodynamics in coupled-cavity
systems provides insight about the behavior of strongly
interacting photons and atoms via a variety of interaction schemes
\cite{hartmann1}. With the capability of tunable coupling and
measurement of individual cavity fields, Coupled-cavity QED could
serve as an useful tool to address the control of quantum
many-body phenomena \cite{review1, review2}. The
Jaynes-Cummings-Hubbard (JCH) model corresponds to a fundamental
configuration exhibiting the quantum phase transition of light
\cite{JCH0,JCH1,JCH2,JCH3,JCH4,JCH5,JCH6,JCH7}. In such a model,
single two-level atoms are embedded in each cavity and the dipole
interaction leads to dynamics involving photonic and atomic
degrees of freedom, which is in contrast to the widely studied
Bose-Hubbard model. Recently, the phase diagrams and excitation
spectra of the JCH model have been discussed in literature
\cite{spectrum0,spectrum1,spectrum2,phase_diag}.

In this paper we examine the eigenstates of the JCH Hamiltonian in
the two-excitation subspace. Our focus is placed on the existence of
bound states as well as their features. It is interesting to note
that two repulsive bosonic atoms can form a bound pair in an optical
lattice \cite{zoller,valiente,Juha}. As we shall indicate below, the
JCH Hamiltonian also supports two-polariton bound states when the
photon-atom interaction is sufficiently strong. In particular, the
two polaritons associated with the bound states exhibit a strong
correlation such that they stay close to together in position space.

To begin with, we first specify the system configuration and the
model Hamiltonian. The one-dimensional JCH model consists of a chain
of $N$-coupled single-mode cavities and each cavity contains a
two-level atom. The cavity mode frequency is $\omega_c$ and the
atomic transition frequency is $\omega_a$. The Hamitonian of the
model in the frame rotating with the atomic frequency $\omega_a$ is
given by $(\hbar=1)$:
\begin{eqnarray}\label{H:JCH}
    H &=& \sum_{n=1}^{N}\Delta a_{n}^{\dag}a_{n}
        + J\sum_{n=1}^{N}
        \left(a_{n+1}^{\dag}a_{n}+a_{n}^{\dag}a_{n+1}\right)
        \nonumber \\
        && + g \sum_{n=1}^{N}  \left(a_{n}\sigma_{n}^{+}
        + a_{n}^{\dag}\sigma_{n}^{-}\right)
\end{eqnarray}
where $\Delta \equiv \omega_c - \omega_a$ is the detuning. The
$a_{n}$ and $a_{n}^{\dag}$ are the annihilation and creation
operators for photons of the $n$-th cavity. The $\sigma_{n}^{\pm}$
are Pauli ladder operators for the atom in the $n$-th cavity. The
$J$ describes the photon tunneling rate between neighboring
cavities, and $g$ is the vacuum Rabi frequency which characterizes
the photon-atom interaction strength. In this paper we assume the
periodic boundary condition such that the cavity labeled by $n=N+1$
corresponds to the cavity $n=1$.

Defining the atomic and photonic excitation number operators as
$\hat{N}_A \equiv \sum_{n=1}^{N} \sigma_{n}^{+}\sigma_{n}^{-}$ and
$\hat{N}_{\gamma} \equiv \sum_{n=1}^{N}a_n^{\dag}a_n$, it is easy to
check that the excitation number (or the polariton number) is a
conserved quantity, i.e., $[H,\hat{N}_A+\hat{N}_{\gamma}]=0$. In
this paper we will focus on states $|\psi \rangle $ with two
excitations only, i.e., $(\hat{N}_A+\hat{N}_{\gamma}) |\psi \rangle
= 2 |\psi \rangle$.

In order to exploit the translational invariance, we define the
following operators via discrete Fourier transform:
\begin{eqnarray}
  \label{def:bk}
    b_{k} &=&
    \frac{1}{\sqrt{N}} \sum_{n=1}^{N}e^{-\frac{2\pi i
    k n}{N}}a_{n} \\
  \label{def:sk}
    s_{k} &=&
    \frac{1}{\sqrt{N}}\sum_{n=1}^{N}e^{-\frac{2\pi i
    k n}{N}}\sigma_{n}^{-}
\end{eqnarray}
where $k=0,1,2,...,N-1$ is related to the (discrete) quasi-momentum
of the excitation. The commutation relations for the operators are
given by:
\begin{eqnarray}
    \left[b_{k},b_{j}^{\dag}\right] &=& \delta_{k j} \\
    \left[s_{k},s_{j}^{\dag}\right] &=&
    -\frac{1}{N} \sum_{n=1}^{N}e^{\frac{2\pi
    i}{N}n(j-k)}\sigma_{n}^z
\end{eqnarray}
where $\sigma_{n}^z$ is the Pauli-Z matrix for the $n$-th atom.
The Hamiltonian in term of operators defined in (\ref{def:bk}) and
(\ref{def:sk}) is partially diagonalized:
\begin{equation}
    H =
        \sum_
        {k=0}^{N-1}\Omega_k b_{k}^{\dag}b_{k}+
        g \sum_{k=0}^{N-1} \left(b_{k}s_{k}^{\dag}
        + b_{k}^{\dag}s_{k}\right)
\end{equation}
where $\Omega_k = \Delta + 2J\cos\frac{2\pi k}{N}$ are normal mode
frequencies (shifted by $\omega_a$) of the field in the
coupled-cavity system in the absence of atoms.  For later purpose,
the ground state of the system is denoted by $|\Phi_0 \rangle$
which is the state with vacuum cavity fields and all atoms being
in their ground levels.

The two-polariton subspace can be spanned by the kets:
$|k\,j\rangle_F \equiv {b_k^{\dag}b_j^{\dag}} |\Phi_0 \rangle$,
$|k\,j\rangle_A \equiv s_k^{\dag}s_j^{\dag}| \Phi_0 \rangle$, and
$|k\rangle_F|j\rangle_A \equiv b_k^{\dag}s_j^{\dag}|\Phi_0 \rangle$
with $k$ and $j$ ranging from 0 to $N-1$. The subscripts $F$ and $A$
are used to denote the field and atomic excitations respectively.
For convenience, $|k\,j\rangle_F$ and $|k\,j\rangle_A$ are not
normalized. Note that $|k\,j\rangle_A$ are generally not orthogonal
to each other because $\langle k '\,j'\vert k\,j\rangle_A
=\delta_{kk'}\delta_{jj'}+\delta_{kj}
\delta_{kk'}\delta_{kj'}-\frac{2}{N}\delta_{k+j,k'+j'}$.
Furthermore, it can be shown that
\begin{eqnarray}
  \label{eq_bs1}
&& b_l s_l^{\dag} |k\,j\rangle_F
        = \delta_{lj}|k\rangle_F|j\rangle_A
        +\delta_{lk}|j\rangle_F|k\rangle_A \\
  \label{eq_bs2}
&& (b_l^{\dag}s_l + b_l s_l^{\dag})|k \rangle_F |j\rangle_A
        = \delta_{lj}|k j \rangle_F
        +\delta_{lk}|k j\rangle_A \\
        \label{eq_bs3}
&& b_l^{\dag}s_l |k\,j\rangle_A
        = \delta_{kl}|k\rangle_F|j\rangle_A
        +\delta_{jl}|j\rangle_F|k\rangle_A \nonumber \\
       && \ \ \ \ \ \ \ \ \ \ \ \ \ \ \
       -\frac{2}{N}|l\rangle_F|[k+j-l]\rangle_A
\end{eqnarray}
with $[x] \equiv x \pmod{N}$. Eqs. (\ref{eq_bs1}-\ref{eq_bs3})
imply that when $H$ operates on $|k\,j\rangle_F$,
$|k\rangle_F|j\rangle_A$ or $|k\,j\rangle_A$, the quantum number
$P \equiv k+j \pmod{N} $ remains unchanged. We may call this as a
conservation of quasi-momentum which is the key to construct
eigenvectors.

A general two-polariton eigenvector with a given quasi-momentum
quantum number $P$ is given by:
\begin{eqnarray}
    |\Psi_P \rangle
        &=& \sum_{(k,j) \in S_P} (
       \alpha_{kj}|kj\rangle_F+
        \beta_{kj}|k\rangle_F|j\rangle_A \nonumber \\
        && \ \ \ +
        \beta_{kj}'|j\rangle_F|k\rangle_A
        +\gamma_{kj}|kj\rangle_A )
\end{eqnarray}
where $S_P$ denotes the set of $(k,j)$ satisfying $k+j\equiv P
\pmod{N}$ and $j \ge k$. To avoid double counting, we set $\beta
_{kk}'=0$. Next by the Schr\"odinger equation, assuming $\lambda$ is
the eigenvalue, $H|\Psi_{P}\rangle = \lambda |\Psi_{P}\rangle$, we
have, for $j > k$,
\begin{eqnarray}
\label{akj}    \lambda\alpha_{kj} &=& \left(\Omega_k+\Omega_j\right)
   \alpha_{kj}
        +g \left(\beta_{kj}+ \beta_{kj}' \right) \\
\label{bkj}    \lambda \beta_{kj} &=& g\alpha_{kj}+\Omega_k
\beta_{kj}+g
    \gamma_{kj}-\frac{2g}{N}
        \sum_{ S_P}\gamma_{k'j'}
        \\
\label{bjk}    \lambda \beta_{kj}' &=& g\alpha_{kj}+\Omega_j
\beta_{kj}'+g
    \gamma_{kj}-\frac{2g}{N}
       \sum_{S_P}\gamma_{k'j'}
        \\
\label{ckj}    \lambda \gamma_{kj} &=& g \left(\beta_{kj}+
\beta_{kj}'\right)
\end{eqnarray}
and for $j=k$,
\begin{eqnarray}
    \lambda \alpha_{kk} &=& 2\Omega_k \alpha_{kk} + g \beta_{kk} \\
    \lambda \beta_{kk} &=& 2g \alpha_{kk}+\Omega_k \beta_{kk}+
        2g \gamma_{kk}-\frac{2g}{N}
       \sum_{ S_P}\gamma_{k'j'}
        \\
    \lambda \gamma_{kk} &=& g \beta_{kk}.
\end{eqnarray}
It is worth noting that the case of even $N$ and odd $P$ is simpler
because this excludes the possibility $k=j$.

We may cast Eqs. (\ref{akj}-17) in the matrix form and solve the
eigensystem directly by standard numerical packages. Since we are
interested in systems with a large $N \gg 1 $, the corresponding
eigenvectors are expected to be insensitive to parity of $N$ and $P$
and this has been verified in our numerical calculations. To
facilitate our discussions, we shall focus on the case of even $N$
and odd $P$ so that Eq. (15-17) are not needed, and Eqs. (11-14)
become:
\begin{equation}\label{eigen_eq}
    {\bf H}_P {\bf V} = \lambda {\bf V}
\end{equation}
where ${\bf H}_P $ is a $2N \times 2N$ matrix:
\begin{equation}\label{H_P}
    {\bf H}_{P} =
    \begin{pmatrix}
     {\bf h}_{k_1,j_1} & {\bf w}_{0} & {\bf w}_{0}
        & \cdots & {\bf w}_{0} \\
     {\bf w}_{0} & {\bf h}_{k_2,j_2} & {\bf w}_{0}
        & \cdots & {\bf w}_{0} \\
     {\bf w}_{0} & {\bf w}_{0} & {\bf h}_{k_3,j_3}
        & \cdots & {\bf w}_{0} \\
     \vdots & \vdots & \vdots & \ddots & \vdots \\
     {\bf w}_{0} & {\bf w}_{0} & {\bf w}_{0} & \cdots
        & {\bf h}_{k_{N/2},j_{N/2}} \\
    \end{pmatrix}
\end{equation}
where $(k_i, j_i) \in S_P$, and ${\bf h}_{k_i,j_i}$ and ${\bf w}_0$
are $4 \times 4$ submatrices:
\begin{eqnarray}
    {\bf h}_{k,j} &=&
    \begin{pmatrix}
     \Omega_k+\Omega_j & g & g & 0 \\
     g & \Omega_k & 0 & g-\frac{2g}{N} \\
     g & 0 & \Omega_j & g-\frac{2g}{N} \\
     0 & g & g & 0 \\
    \end{pmatrix}
\\
    {\bf w}_{0} &=&
    \begin{pmatrix}
     0 & 0 & 0 & 0 \\
     0 & 0 & 0 & -\frac{2g}{N} \\
     0 & 0 & 0 & -\frac{2g}{N} \\
     0 & 0 & 0 & 0 \\
    \end{pmatrix}
\end{eqnarray}
In this way the eigenvector ${\bf V}$ takes the form:
\begin{equation}
    {\bf V} =  \left({
        {\bf v}_{k_1,j_1}, {\bf v}_{k_2,j_2}, \cdots ,
        {\bf v}_{k_{N/2},j_{N/2}} }\right)^T
\end{equation}
with ${\bf v}_{k,j} = \left({
       \alpha_{kj},  \beta_{kj},  \beta_{kj}' ,
        \gamma_{kj} }\right) ^T$.
We remark that ${\bf H}_{P}$ is not symmetric and this is due to
the fact that non-orthogonal atomic basis vectors $| k j
\rangle_A$ have been used to express the eigenvectors.

\begin{figure}[tbp]
\includegraphics [width=8.5 cm] {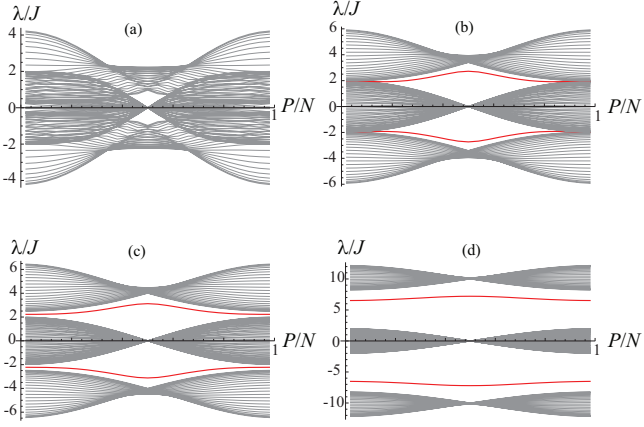}
\caption{(Color online) Eigenvalues of the $H$ for $N=50$ and odd
$P$ and $\Delta=0$ for (a) $g/J=0.1$, (b) $g/J=1.7$, (c) $g/J=2$ and
(d) $g/J=5$. The eigenvalues are joint by lines for ease of
visualization.} \label{fig:lc}
\end{figure}

In Fig. 1, we illustrate the eigenvalues of $H$ as a function of $P$
at resonance $(\Delta =0)$ for various photon-atom interaction
strengths. We find that if $g/J$ is sufficiently large (Fig. 1c and
1d), then there exist two branches of discrete eigenvalues (marked
in red). The discrete eigenvalues correspond to bound states in the
large $N$ limit and they are isolated from the quasi-continuous
bands of eigenvalues. By lowering the photon-atom interaction (Fig.
1b), the bands and the discrete eigenvalues become closer, and
eventually the discrete eigenvalues disappear in the bands if $g$ is
too small (Fig. 1a). We will estimate the critical value of $g$ for
the occurrence of bound states near the end of this paper.

Let $|\lambda_b \rangle$ be the normalized eigenvector with the
discrete eigenvalue $\lambda_b$. Although the notion bound states
is for infinite $N$ systems, the confined nature of $|\lambda_b
\rangle$ starts to emerge in real space at finite large $N$. This
is illustrated in Fig. 2 in which the probability distribution of
excitations in real space are plotted. Specifically, we calculate
the joint probabilities defined by,
\begin{eqnarray}
&& p_{nm}^{FF} = | \langle \lambda_b | \frac {a_n^{\dag} a_m^{\dag}}
{\sqrt{1+\delta_{nm}}} |\Phi_0 \rangle |^2 \\
&& p_{nm}^{AA} = | \langle \lambda_b | \sigma_n^+ \sigma_m^+
    |\Phi_0 \rangle |^2 \\
&& p_{nm}^{FA} = | \langle \lambda_b |  {a_n^{\dag} \sigma_m^+}
 |\Phi_0 \rangle |^2
\end{eqnarray}
Here $m$ and $n$ are indices for the cavity positions, and hence
$p_{nm}^{FF}$ ($p_{nm}^{AA}$) is the joint probability of having
photonic (atomic) excitations in the $n$-th and $m$-th cavity.
Similarly, $p_{nm}^{FA}$ is the joint probability of a single photon
in $n$-th cavity and an excited atom in the $m$-th cavity. For a
given value of $P$, all the three joint probabilities depend on
$n-m$ only.

\begin{figure}[tbp]
\includegraphics [width=8 cm] {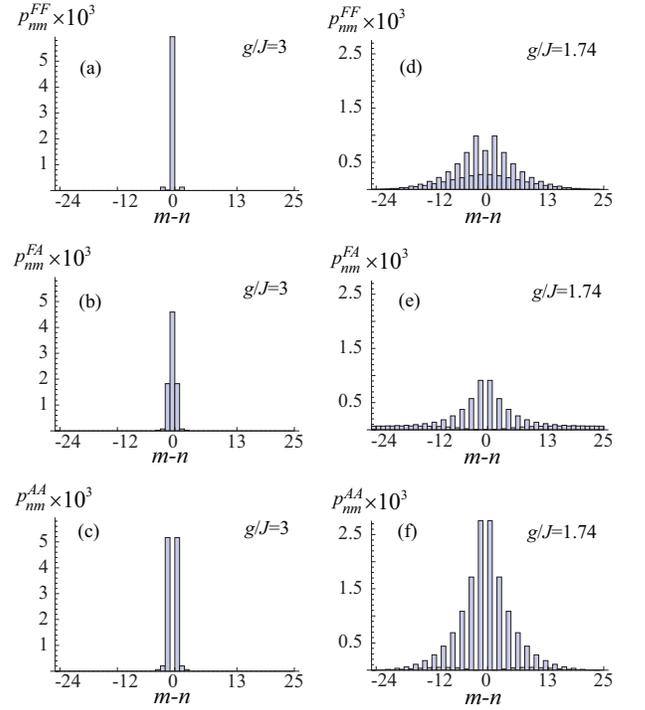}
\caption{(Color online) Joint probabilities distribution associated
with bound states for the system with $\Delta=0$, $P=1$ and $N=50$
at various coupling strengths. The right column corresponds to $g$
slightly above the critical value.} \label{fig:abc}
\end{figure}

In Fig. 2(a-c), we see that the joint probabilities are strongly
localized around $m=n$ when the interaction strength $g$ is
sufficiently away from the critical value ($g_c \approx \sqrt 3 J)$.
Note that $p_{nm}^{AA}$ vanishes at $n=m$ because $\sigma_n^{+2}=0$,
but $p_{nm}^{AA}$ peaks at $n=m\pm 1$. The localized feature means
that the two polaritons are spatially confined together and behave
like a composite object. As the photon-atom interaction $g$
decreases, the binding becomes weaker. This is shown in the wider
distributions in Fig. 2(d-e) in which $g/J$ is slightly above the
critical ratio. We remark that although Fig. 2 is plotted with $P=1$
as an illustration, our numerical results indicate that the
localized feature of bound states appear for the whole range of $P$.
For example, in the case $N=50$ and $g/J =5$, by adding up the same
site probabilities and the nearest neighbor probabilities, the sum
is over $98.5 \%$ for $0 \le P \le N-1$.

Although an exact analytic solution of the bound states is not
available, some analytic insights may be obtained in certain
limiting cases. From Eqs. (\ref{akj}-\ref{ckj}), non-zero
eigenvalues $\lambda$ are the roots of an algebraic equation
$G(\lambda)=0$, where $G$ is defined by,
\begin{equation}\label{lambda_eq}
    G(\lambda) = 1-\frac{2g^2}{N}
      \sum_{(k,j)\in S_P}
        \frac{C_{kj}(\lambda)}
            {(1+\delta_{kj})D_{kj}(\lambda)}
\end{equation}
with
\begin{eqnarray}
&& C_{kj}(\lambda) = \left(2\lambda-\Omega_k-\Omega_j\right)
            \left(\lambda-\Omega_k-\Omega_j\right)
\\
&& D_{kj}(\lambda) =
g^2\left(2\lambda-\Omega_k-\Omega_j\right)^2 \nonumber \\
&& \ \ \ \ \ \ \ \ \ \ \ \ -\lambda
\left(\lambda-\Omega_k\right)\left(\lambda-\Omega_j\right)
            \left(\lambda-\Omega_k-\Omega_j\right).
\end{eqnarray}
In the strong coupling limit with $g \gg J$, we find that the
eigenvalues $\lambda_b$ for bound states are approximately given by,
\begin{equation}\label{lb}
    \lambda_b \approx \pm\sqrt{2}\left[g-
        \frac{J^2}{2g}\left(4+5\cos\frac{2\pi P}{N}\right)\right].
\end{equation}
This is obtained by noting that at $J=0$ (i.e., when the cavities
are decoupled), the $\pm \sqrt 2 g$ are dressed energies known in
the resonance Jaynes-Cummings model. Treating this as a zeroth order
approximation, we can make a Taylor expansion of Eq.
(\ref{lambda_eq}) in a power series $J/g$. Eq. (\ref{lb}) is then
obtained as an approximation solution of $\lambda_b$ up to the first
order in $J/g$.

\begin{figure}[tbp]
\includegraphics [width=7 cm] {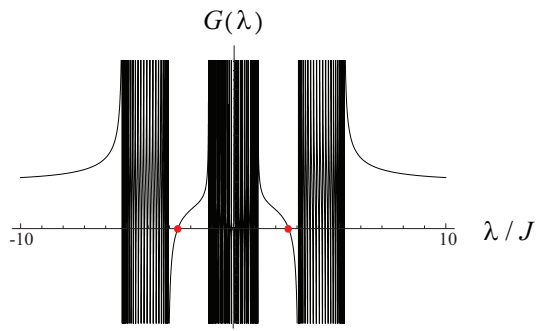}
\caption{(Color online) An illustration of the function $G$ for the
system with $\Delta=0$, $N=50$, $P=15$ and $g/J=2$. Non-zero
eigenvalues of the Hamiltonian are roots of $G(\lambda) =0$. The red
dots inside the gaps are roots corresponding to bound states.}
\label{fig:lc}
\end{figure}

In the $N \gg 1$ limit, Eq. (\ref{lambda_eq}) can also be used to
estimate the critical value of $g$ below which bound states cease to
exist. This is done by observing that the function $G(\lambda)$ form
bands, and $\lambda_b$'s exist in the gaps between the bands (Fig.
3). As $g$ decreases the gaps between bands become narrower. At a
critical $g=g_c$, the bands start overlap and no bound states are
supported for $g<g_c$. Therefore $g_c$ can be estimated by
determining when the band edges overlap. In the $N \gg 1$ limit, the
bands are filled up by those $\lambda$'s that give zero denominators
in (\ref{lambda_eq}) [i.e., $G(\lambda) \to \pm \infty$] as shown in
Fig. 3, and hence the band edges can be estimated. Base on this
estimation scheme, the critical value $g_c$ in the $N \to \infty$
limit is plotted as a function of $P$ (Fig. 4). Note that $g_c$ is a
maximum at $P=0$ and $P=N-1$ and it equals $\sqrt 3 J$.


\begin{figure}[tbp]
\includegraphics [width=6.5 cm] {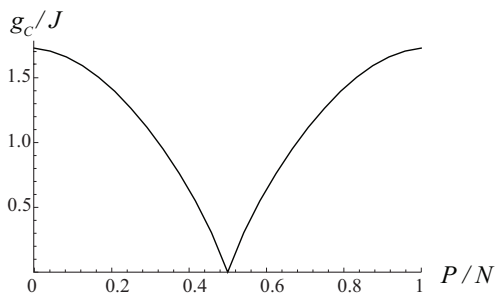}
\caption{The critical ratio $g_c/J$ as a function of $P$ for
$\Delta=0$ systems.} \label{fig:lc}
\end{figure}

To conclude, we have addressed the spectrum as well as the
conditions for the existence of two-polariton bound states of the
JCH model. These bound states are composite objects characterized by
the quasi-momentum quantum number $P$ and  they process a strong
spatial correlation. In view of the previously studied bound atom
pairs in Bose-Hubbard model \cite{zoller,valiente,Juha}, our work is
a generalization to bound polariton pairs in a coupled cavity QED
system. Although our analysis have been confined to the zero
detuning case, bound states are also found for detuned systems
according to our numerical solutions of Eq. (18). The effect of
detuning would control the nature of polaritons such that the two
excitations can be mainly photonic (atomic) at positive (negative)
large $\Delta$.  We also remark that the dynamics of a single
polariton in the JCH model has been discussed recently
\cite{time_evo}, and we expect the two-polariton problem may have
richer dynamical features because bound states would enable
interesting correlated two-polariton transport. We hope to address
this issue in the future.

\begin{acknowledgments}
This work is supported in part by the Research Grants Council of the
Hong Kong Special Administrative Region, China (Project No. 401408).
\end{acknowledgments}

\end{document}